\documentclass[12pt,a4paper]{article}
\usepackage{amsmath,amssymb}
\usepackage{epsfig}
\newcommand{\No}{N}
\begin{document}

\begin{center}
\Large\textbf{PROBLEM OF COSMOLOGICAL SINGULARITY, INFLATIONARY
COSMOLOGY AND GAUGE THEORIES OF GRAVITATION
}\\
\bigskip
\normalsize A. V. Minkevich\\
\bigskip
 \textit{\small $^1$Department of Theoretical Physics, Belarussian State University,\\
av. F. Skoriny 4, 220050, Minsk, Belarus,\\
 $^2$Department of Physics and Computer Methods, Warmia and Mazury University in Olsztyn,
 Poland\\
\normalsize  E-mail: MinkAV@bsu.by; awm@matman.uwm.edu.pl}
\end{center}
\medskip
\begin{flushright}
\begin{minipage}{\textwidth}\small
\textbf{Abstract.} Problem of cosmological singularity is
discussed in the framework of gauge theories of gravitation. Generalizing
cosmological Friedmann equations (GCFE) for homogeneous isotropic models including
scalar fields and usual gravitating matter are introduced. It is shown that by
certain restrictions on equation of state of gravitating matter and indefinite
parameter of GCFE  generic feature of inflationary cosmological models of flat, open
and closed type is their regular bouncing character.
\end{minipage}
\end{flushright}

\section{Introduction}

Problem of cosmological singularity (PCS) is one of the most principal problems of
general relativity theory (GR) and relativistic cosmology. The appearance of
singular state with divergent values of energy density and invariants of curvature
tensor in cosmological solutions of GR, which limits the existence of the universe
in the past, is inevitable according to Hawking-Penrose theorems, if gravitating
matter satisfies so-called energy dominance conditions [1, 2]. There were many
attempts to resolve PCS in the frame of GR as well as various generalizations of
Einstein's theory of gravitation. The radical idea of quantum birth of the Universe
was introduced in order to avoid the PCS \cite{mc3}. A number of regular
cosmological solutions was obtained in the frame of metric theories of gravitation
and also other theories, in the frame of which gravitation is described by using
more general geometry than the Riemannian one (see [4--8]  and refs given herein).
However, still now we do not have the resolution of PCS. Really, the resolution of
PCS means not only containing regular cosmological solutions, but also excluding
singular solutions of cosmological equations by using physically reasonable initial
conditions. Note that gravitation theory and cosmological equations have to satisfy
the correspondence principle with Newton's theory of gravitation and GR in the case
of usual gravitating systems with sufficiently small energy densities and weak
gravitational fields excluding nonphysical solutions. So metric theories of
gravitation based on gravitational Lagrangians including terms quadratic in the
curvature tensor lead to cosmological equations with high derivatives, and although
these theories permit to obtain regular cosmological solutions with Friedmann and de
Sitter asymptotics, however, they possess nonphysical solutions also. At last time
some regular cosmological solutions were found in the frame of superstring theory
(brane cosmology), but these solutions also do not resolve the PCS (see for example
[7--10]).

Present paper is devoted to study the PCS in the frame of gauge theories of
gravitation (GTG), at first of all of the Poincare GTG and metric-affine GTG. Note
that GTG are natural generalization of GR by applying the local gauge invariance
principle, which is a base of modern theory of fundamental physical interactions
(see review \cite{Hehl}). The first attempt to apply the simplest Poincare GTG - the
Einstein-Cartan theory --- in order to resolve PCS was made in Refs
\cite{mc12,mc13}, where some nonsingular models with spinning matter were built.
Note that these results depend essentially on classical model of spinning matter,
and in the case of usual spinless matter the Einstein-Cartan theory is identical to
GR and hence singular Friedmann models of GR are its exact solutions. The next step
to investigate the PCS in the frame of Poincare GTG was made in Ref.\cite{mc14} by
using sufficiently general gravitational Lagrangian including both a scalar
curvature and terms quadratic in the curvature and torsion tensors. Note that in the
frame of GTG quadratic in the curvature tensor terms of gravitational Lagrangian do
not lead to high derivatives in cosmological equations for homogeneous isotropic
models modifying Friedmann cosmological equations of GR (see below). The conclusion
about possible existence of limiting energy (mass) density for usual spinless
gravitating matter was obtained from deduced cosmological equations leading to
regular in metrics cosmological solutions. (Later the possible existence of limiting
energy density was discussed in Ref. \cite{mc15} by modifying cosmological Friedmann
equations of GR.) Because cosmological equations of Poincare GTG are valid in the
frame of the most general GTG --- metric-affine GTG \cite{mc16,mc17}, the same
conclusions are valid also in metric-affine GTG . Further study of homogeneous
isotropic models in GTG has showed that these theories possess important
regularizing properties and lead to gravitational repulsion effect nearby limiting
energy density, which can be caused by different physical factors [18--21]. Because
the behaviour of cosmological models at the beginning of cosmological expansion in
GTG depends essentially on equation of state of gravitating matter, we have to know
this equation for gravitating matter at extreme conditions in order to build more
realistic cosmological models. Unified gauge theories of strong and electroweak
interactions with spontaneous symmetry breaking are the modern base of matter
description at extreme conditions. Inflationary cosmology as important part of the
theory of early Universe was built by using gauge theories of elementary particles
\cite{mc22,mc23}. A number of problems of standard Friedmann cosmology were resolved in the
frame of inflationary cosmology. Most inflationary cosmological models discussed in
literature are singular and their study is given from Planckian time. Essential
contribution to energy density in inflationary models is given by so-called
gravitating vacuum (connected with scalar fields), for which pressure $p$ and energy
density $\rho>0$ are connected in the following way $p=-\rho$.  As it was shown in
Refs. \cite{mc18,mc24}, in the frame of GTG gravitating vacuum with sufficiently
large energy density can lead to the vacuum gravitational repulsion effect (VGRE) in
the case of systems including also usual gravitating matter, that allows to build
regular inflationary cosmological models. At first the VGRE was discussed in the
frame of Poincare GTG \cite{mc18} in the case of homogeneous isotropic models
including radiation and gravitating vacuum with $\rho=\mathrm{const}$. Regularizing
role of gravitating vacuum in GTG was analyzed in Refs [25--28].  Unlike GR where
gravitating vacuum can lead to a bounce only in the case of closed cosmological
models \cite{mc29,mc30}, in GTG the VGRE allows to build bouncing inflationary
models of flat, open and closed type. In order to build more realistic inflationary
models, homogeneous isotropic models including scalar fields and ultrarelativistic
matter were analyzed by applying cosmological equations of GTG \cite{mc31}. It was
shown that by certain conditions GTG permit to build regular inflationary
cosmological models of flat, open and closed type with dominating ultrarelativistic
matter at a bounce, and the greatest part of cosmological solutions have bouncing
character. By taking into account that regular in metrics cosmological solutions in
GTG take place in the case of gravitating matter with $p\neq
\frac{1}{3}\rho$, we will study below homogeneous isotropic
models including  besides scalar fields and ultrarelativistic matter also
gravitating matter with $p\neq \frac{1}{3}\rho$. In Section 2  cosmological
equations of GTG describing such models are introduced. In Section 3 most important
general solutions properties of introduced cosmological equations are studied. In
Section 4 bouncing character of inflationary cosmological models in GTG is analyzed.

\section{Generalized cosmological Friedmann equations in GTG}

Homogeneous isotropic models in GTG are described by the following generalized cosmological
Friedmann equations (GCFE)
\begin{equation}
\frac{k}{R^2}+\left\{\frac{d}{dt}\ln\left[R\sqrt{\left|1-\beta\left(\rho-
3p\right)\right|}\,\right]\right\}^2=\frac{8\pi}{3M_p^2}\,\frac{\rho-
\frac{\beta}{4}\left(\rho-3p\right)^2}{1-\beta\left(\rho-3p\right)}
\, ,
\end{equation}
\begin{equation}
\frac{\left[\dot{R}+R\left(\ln\sqrt{\left|1-\beta\left(\rho-
3p\right)\right|}\,\right)^{\cdot}\right]^\cdot}{R}=
-\frac{4\pi}{3M_p^2}\,\frac{\rho+3p+\frac{\beta}{2}\left(\rho-3p\right)^2}{
1-\beta\left(\rho-3p\right)}\, ,
\end{equation}
where $R(t)$ is the scale factor of Robertson-Walker metrics, $k=+1,0,-1$ for
closed, flat, open models respectively, $\beta$  is indefinite parameter with
inverse dimension of energy density, $M_p$  is Planckian mass, a dot denotes
differentiation with respect to time\footnote{Parameter $\beta$ is defined as
$\beta=-\frac{1}{3}\left(16\pi\right)^2f\,M_p^{-4}$, where $f$ is linear combination
of coefficients at terms of gravitational Lagrangian quadratic in the curvature
tensor.}. (The system of units with $\hbar=c=1$ is used). Eqs.(1)--(2) are identical
to Friedmann cosmological equations of GR if $\beta$=0. At first the GCFE were
deduced in Poincare GTG
\cite{mc14}, and later it was shown that Eqs.(1)--(2) take place also in
metric-affine GTG
\cite{mc16,mc17}. From Eqs. (1)--(2) follows the conservation law in usual form
\begin{equation}
\dot{\rho}+3H\left(\rho+p\right)=0,
\end{equation}
where $H=\frac{\dot{R}}{R}$   is the Hubble parameter. Besides cosmological
equations (1)--(2) gravitational equations of GTG lead to the following relation for
torsion function $S$ and nonmetricity function $Q$
\begin{equation}
S-\frac{1}{4}Q=-\frac{1}{4}\,\frac{d}{dt}
\ln\left|1-\beta(\rho-3p)\right|.
\end{equation}
In Poincare GTG $Q=0$ and Eq. (4) determines the torsion function. In metric-affine
GTG there are three kinds of models \cite{mc17}: in the Riemann-Cartan space-time
($Q=0$), in the Weyl space-time ($S=0$), in the Weyl-Cartan space-time ($S\neq 0$,
$Q\neq 0$, the function $S$ is proportional to the function $Q$). The value of
$|\beta|^{-1}$ determines the scale of extremely high energy densities. Classical
description of gravitational field is valid, if we suppose that $|\beta|^{-1}<1
M_p^4$.  The GCFE (1)--(2) coincide practically with Friedmann cosmological
equations of GR if the energy density is small $\left|\beta(\rho-3p)\right|\ll 1$
($p\neq\frac{1}{3}\rho$). The difference between GR and GTG can be essential at
extremely high energy densities $\left|\beta(\rho-3p)\right|\gtrsim 1$.
Ultrarelativistic matter with equation of state $p=\frac{1}{3}\rho$ is exceptional
system because Eqs. (1)--(2) are identical to Friedmann cosmological equations of GR
in this case independently on values of energy density. The behaviour of solutions
of Eqs. (1)--(2) depends essentially on equation of state of gravitating matter at
extreme conditions and on sign of parameter $\beta$. By given equation of state
$p=p(\rho)$  the GCFE (1)--(2) lead to regular in metrics cosmological solutions in
the following cases: 1) $\beta>0$ and $p<\frac{1}{3}\rho$, 2) $\beta<0$ and
$p>\frac{1}{3}\rho$  \cite{mc4}. The investigation of models including scalar fields
on the base of Eqs (1)--(2) shows that the choice $\beta<0$ permits to exclude the
divergence of time derivative of scalar fields \cite{mc31}. In connection with this
we put below that parameter $\beta$ is negative and $|\beta|^{-1} < 1\cdot M_p^4$.

In the case of gravitating vacuum with constant energy
density $\rho_v=\mathrm{const}>0$ the GCFE (1)--(2) are reduced to Friedmann
cosmological equations of GR and $S=Q=0$, this means that de Sitter solutions for
metrics with vanishing torsion and nonmetricity are exact solutions of GTG
\cite{mc32} and hence inflationary models can be built in the frame of GTG.

In order to analyze inflationary cosmological models in GTG let us
consider systems including scalar field   minimally coupled with
gravitation and gravitating matter in the form of several
components with energy densities $\rho_i$  and pressures $p_i=w_i
\rho_i$, where $w_i =\mathrm{const}$ ($i$ is a number of
component). Later we will consider two components:
ultrarelativistic matter with energy density $\rho_r$  and
$w_r=\frac{1}{3}$, and gravitating matter with $w>\frac{1}{3}$.
 If the interaction between scalar field and all components of
gravitating matter is negligible, the energy density $\rho$  and
pressure $p$   take the form
\begin{equation}
\rho=\frac{1}{2}\dot{\phi}^2+V(\phi)+\sum\rho_i, \qquad
p=\frac{1}{2}\dot{\phi}^2-V(\phi)+\sum w_i \rho_i,
\end{equation}
where the summation is taken over all components.
 The conservation law (3) leads to the scalar field equation
\begin{equation}
\ddot{\phi}+3H\dot{\phi}=-\frac{dV}{d\phi},
\end{equation}
where $V$ is scalar field potential and to the conservation laws
for all components of gravitating matter
\begin{equation}
\dot{\rho}_i+3H\rho_i (1+w_i)=0.
\end{equation}
Eqs (7) have integrals in the following form
\begin{equation}
\rho_i R^{3(1+w_i)}={\rm const}.
\end{equation}
By using Eqs.(5)--(7) the GCFE (1)--(2) can be transformed to the following form
\begin{eqnarray}
& & \frac{k}{R^2}
Z^2+\left\{H\left[1-2\beta(2V+\dot{\phi}^2)-\frac{1}{2}\beta\sum
\rho_i\left(9w_i^2-1\right)\right]-3\beta
V'\dot{\phi}\right\}^2\nonumber \\
& & \phantom{+H\left[1-2\beta\right]} =\frac{8
\pi}{3M_p^2}\,\left\{\sum\rho_i+
\frac{1}{2}\dot{\phi}^2+V-\frac{1}{4}\beta\left[4V-\dot{\phi}^2-
\sum \rho_i\left(3w_i-1\right)
\right]^2\right\}Z,\\
&
&\dot{H}\left\{1-2\beta\left[2V+\dot{\phi}^2+\frac{1}{4}\sum\rho_i\left(9w_i^2-1\right)
\right]\right\}Z\nonumber\\
& & \phantom{H}+H^2\left\{\left[
1-4\beta(V-4\dot{\phi}^2)-4\beta\sum\rho_i\left(1-\frac{9}{8}w_i
-\frac{9}{2}w_i^2-\frac{27}{8}w_i^3\right)\right]Z\right.\nonumber\\
& & \phantom{H}\left. -\frac{9}{2}\beta^2\left[\sum\rho_i
\left(1-2w_i-3w_i^2\right)-2\dot{\phi}^2\right]^2\right\}\nonumber\\
& &\phantom{H} +12\beta
H\dot{\phi}V'\left\{1-2\beta\left[2V+\dot{\phi}^2+\frac{1}{4}\sum\rho_i(9w_i^2-1)\right]
\right\}\nonumber\\
& &\phantom{H}-
3\beta\left[(V''\dot{\phi}^2-V'{}^2)Z+6\beta\dot{\phi}^2
V'{}^2\right]
\nonumber\\
& &\phantom{\dot{H}\left[\right]} =\frac{8
\pi}{3M_p^2}\,\left\{V-\dot{\phi}^2-\frac{1}{2}\sum\rho_i (1+3w_i)
-\frac{1}{4}\beta\left[4V-\dot{\phi}^2-\sum\rho_i(3w_i-1)\right]^2\right\}Z,\qquad
\end{eqnarray}
where $Z=1-\beta\left[4V-\dot{\phi}^2-\sum\rho_i(3w_i-1)\right]$,
$\displaystyle V'=\frac{dV}{d\phi}$, $\displaystyle
V''=\frac{d^2V}{d\phi^2}$. Relation (4) takes the form
\begin{equation}
S-\frac{1}{4}Q=\frac{3\beta}{2}\,\frac{H\left[\dot{\phi}^2+\frac{1}{2}\sum\rho_i
\left(3w_i^2+2w_i-1\right)\right]+V'\dot{\phi}}{Z}.
\end{equation}
Unlike GR the cosmological equation (9) leads to essential
restrictions on admissible values of scalar field and gravitating
matter. Imposing $\beta<0$, we obtain from Eq. (9) in the case
$k=0,+1$
\begin{equation}
Z\geq 0 \qquad \text{or}\qquad \dot{\phi}^2\leq
4V+|\beta|^{-1}-\sum\rho_i\left(3w_i-1\right).
\end{equation}
Inequality (12) is valid also for open models discussed below. The
region $\Sigma$ of admissible values of scalar field  $\phi$, time
derivative $\dot{\phi}$ and energy densities $\rho_i$ (excluding
$\rho_r$) in space $P$ of these variables determined by (12) is
limited by bounds $L_{\pm}$
\begin{equation}
\dot{\phi}=\pm
\left[4V+|\beta|^{-1}-\sum\rho_i(3w_i-1)\right]^{\frac{1}{2}}.
\end{equation}
From Eq. (9) the Hubble parameter on the bounds $L_{\pm}$  is equal to
\begin{equation}
  H=\frac{3\beta V'\dot{\phi}}{1-2 \beta(2V+\dot{\phi}^2)-\frac{1}{2}\beta\sum\rho_i(9w_i^2-1)}.
\end{equation}
According to (14) the right-hand part of Eq. (11) is equal to $\frac{1}{2}H$, this
means that the torsion (nonmetricity) will be regular, if the Hubble parameter is
regular.  In Sections 3-4 our main attention will be turned to study properties of
solutions of GCFE (9)--(10).

\section{Some general properties of GCFE solutions for inflationary models}

Let us consider the most important general properties of cosmological solutions of
GCFE (9)--(10). At first, note by given initial conditions for scalar field ($\phi$,
$\dot{\phi}$) and values of $R$ and $\rho_i$ there are two different solutions
corresponding to two values of the Hubble parameter following from Eq.~(9):
\begin{equation}
  H_{\pm}=\frac{3\beta V'\dot{\phi}\pm\sqrt{D}}{1-2 \beta(2V+\dot{\phi}^2)-
  \frac{1}{2}\beta\sum\rho_i(9w_i^2-1)},
\end{equation}
where
\begin{equation}
D=\frac{8\pi}{3M_p^2}\,\left\{\sum\rho_i+
\frac{1}{2}\dot{\phi}^2+V-\frac{1}{4}\beta\left[4V-\dot{\phi}^2
-\sum\rho_i\left(3w_i-1\right)\right]^2\right\}Z-\frac{k}{R^2}\,Z^2\ge
0
\end{equation}
Unlike GR, the values of $H_{+}$ and $H_{-}$ in GTG are sign-variable and, hence,
both solutions corresponding to $H_{+}$ and $H_{-}$ can describe the expansion as
well as the compression in dependence on their sign. Below we will call solutions of
GCFE corresponding to $H_{+}$ and $H_{-}$ as $H_{+}$-solutions and $H_{-}$-solutions
respectively. In points of bounds $L_{\pm}$ we have $D=0$, $H_{+}=H_{-}$ and the
Hubble parameter is determined by (14). Eqs. (9)--(10) are satisfied on the bounds
$L_{\pm}$, corresponding solutions of GCFE -- $L_{\pm}$--solutions -- are their
particular solutions; $H_{-}$--solutions reach the bounds $L_{\pm}$ and
$H_{+}$-solutions originate from them (see below). By using Eqs. (15)--(16) it is
easy to shown, that in points of bounds $L_{\pm}$ the derivatives $\dot{H}_{+}$ and
$\dot{H}_{-}$ are equal: $\lim {\dot{H}_{+}}=\lim{\dot{H}_{-}}$ at $Z\to 0$. As
result, we have the smooth transition from $H_{-}$-solution to $H_{+}$-solution on
bounds $L_{\pm}$. At the same time, the value of the time derivative of $\dot{H}$
for $L_{\pm}$-solutions according to (14) is not equal to $\lim{\dot{H}_{\pm}}$ at
$Z\to 0$, and by transition from $H_{-}$-solution to $L_{\pm}$-solutions and from
$L_{\pm}$-solution to $H_{+}$-solution a finite jump of the derivative $\dot{H}$
takes place. Note, that according to Eq. (11) the functions $S$ and $Q$ have the
following asymptotics at $Z\to 0$: $\left|S-\frac{1}{4}Q\right|\sim Z^{-1/2}$ for
$H^{+}$- and $H_{-}$-solutions.

In order to study the behaviour of cosmological models at the beginning of
cosmological expansion, let us analyze extreme points for the scale factor $R(t)$:
$R_0=R(0)$, ${H_0=H(0)=0}$. Denoting values of quantities at $t=0$ by means of index
"0", we obtain from (9)--(10):
\begin{eqnarray}
& &\frac{k}{R_0^2} Z_0^2+9\beta^2
V'{}_0^2\dot{\phi}_0^2\nonumber\\
& &\phantom{\frac{k}{R_0^2} Z_0^2} =\frac{8
\pi}{3M_p^2}\,\left\{\sum\rho_{i0}+\frac{1}{2}\dot{\phi_0}^2
+V_0-\frac{1}{4}\beta\left[4V_0-\dot{\phi}_0^2-\sum\rho_{i0}\left(3w_i-1\right)\right]^2\right\}Z_0,\\
& &\dot{H}_0=\left\{\frac{8
\pi}{3M_p^2}\,\left[V_0-\dot{\phi}_0^2-\frac{1}{2}\sum\rho_{i0}\left(1+3w_i\right)-\frac{1}{4}\beta\left(4V_0
-\dot{\phi}_0^2-\sum\rho_{i0}\left(3w_i-1\right)\right)^2\right]Z_0\right.\nonumber\\
&
&\left.\phantom{H_0=}+3\beta\left[(V''{}_0\dot{\phi}_0^2-V'{}_0^2)Z_0
+6\beta\dot{\phi}_0^2 V'{}_0^2\right]\right\}\nonumber\\
& &\phantom{H_0=\frac{8 \pi}{3M_p^2}\,V_0-\dot{\phi}_0^2}\times
\left\{1-2\beta\left[2V_0+\dot{\phi}_0^2+\frac{1}{4}\sum\rho_{i0}\left(9w_i^2-1\right)\right]
\right\}^{-1}Z_0^{-1},
\end{eqnarray}
where
$Z_0=1-\beta\left[4V_0-\dot{\phi}_0^2-\sum\rho_{i0}\left(3w_i-1\right)\right]$.
A bounce point is described by Eq. (17), if the value of
$\dot{H}_0$ is positive. By using Eq.(17) we can rewrite the
expression of $\dot{H}_0$ in the form
\begin{eqnarray}
& &\dot{H}_0=\left\{\frac{8
\pi}{M_p^2}\,\left[V_0+\frac{1}{2}\sum\rho_{i0}\left(1-w_i\right)-\frac{1}{4}\beta\left(4V_0
-\dot{\phi}_0^2-\sum\rho_{i0}\left(3w_i-1\right)\right)^2\right]\right.\nonumber\\
& &\phantom{H_0=}\left.+3\beta(V''{}_0\dot{\phi}_0^2-V'{}_0^2)
-\frac{2k}{R_0^2}Z_0\right\}
\left\{1-2\beta\left[2V_0+\dot{\phi}_0^2+\frac{1}{4}\sum\rho_{i0}\left(9w_i^2-1\right)\right]
\right\}^{-1}.\qquad
\end{eqnarray}

We see from (19) unlike GR the presence of gravitating matter (with $w_i\le 1$) does
not prevent from the bounce realization\footnote{In GR a bounce is possible only in
closed models if the following condition
$V_0-\dot{\phi}_0^2-\frac{1}{2}\sum\rho_{i0}(1+3w_i)>0$ takes place.}. In the case
of various scalar field potentials applying in inflationary cosmology Eq.(17)
determines in space $P$ so-called "bounce surfaces" depending on parameter $\beta$
and energy density of ultrarelativistic matter parametrically. In the case of closed
and open models families of bounce surfaces depend also on the scale factor $R_0$.By
giving concrete form of potential $V$ and choosing values of $R_0$, $\phi_0$,
$\dot{\phi}_0$ and $\rho_{i0}$ at a bounce, we can obtain numerically particular
bouncing solutions of GCFE for various values of parameter $\beta$.

The analysis of GCFE shows, that properties of cosmological solutions depend
essentially on parameter  $\beta$, i.e. on the scale of extremely high energy
densities. From physical point of view interesting results can be obtained, if the
value of $|\beta|^{-1}$   is much less than the Planckian energy density
\cite{mc33}, i.e. in the case of large in module values of
parameter $\beta$ (by imposing $M_p=1$). In order to investigate
cosmological solutions at the beginning of cosmological expansion
in this case, let us consider the GCFE by supposing that
\begin{eqnarray}
& &
\left|\beta\left[4V-\dot{\phi}^2-\sum\rho_i(3w_i-1)\right]\right|\gg
1,\nonumber\\
& &\sum\rho_i+\frac{1}{2}\dot{\phi}^2+V\ll\left|\beta\right|
\left[4V-\dot{\phi}^2-\sum\rho_i(3w_i-1)\right]^2.
\end{eqnarray}
Note that the second condition (20) does not exclude that ultrarelativistic matter
energy density can dominate at a bounce. We obtain:
\begin{multline}
\frac{k}{R^2}+\frac{\left\{2H\left[2V+
\dot{\phi}^2+\frac{1}{4}\sum\rho_i(9w_i^2-1)\right]+3V'\dot{\phi}\right\}^2}{\left[4V-\dot{\phi}^2
-\sum\rho_i(3w_i-1)\right]^2}\\
 =\frac{2\pi}{3M_p^2}
\left[4V-\dot{\phi}^2-\sum\rho_i(3w_i-1)\right],
\end{multline}
\begin{multline}
  \dot{H}\left[2V+\dot{\phi}^2+\frac{1}{4}\sum\rho_i\left(9w_i^2-1\right)\right]\left[4V-\dot{\phi}^2
  -\sum\rho_i\left(3w_i-1\right)\right]+\\
  H^2\left\{2\left[V-4\dot{\phi}_2+\sum\rho_i\left(1-\frac{9}{8}w_i-
  \frac{9}{2}w_i^2-\frac{27}{8}w_i^3\right)\right]\left[4V-\dot{\phi}^2-\sum\rho_i
  \left(3w_i-1\right)\right]\right.\\
  \left.-\frac{9}{4}\left[2\dot{\phi}^2-\sum\rho_i\left(1-2w_i-3w_i^2\right)\right]^2\right\}
  -12H V'\dot{\phi}\left[2V+\dot{\phi}^2+\frac{1}{4}\sum\rho_i\left(9w_i^2-1\right)\right]\\
  + \frac{3}{2}\left(V''\dot{\phi}^2-V'^2\right)\left[4V-\dot{\phi}^2-\sum\rho_i\left(3w_i-1\right)\right]-
  9V'^2\dot{\phi}^2\\
  =  \frac{\pi}{3M_p^2}\left[4V-\dot{\phi}^2-\sum\rho_i\left(3w_i-1\right)\right]^3.
\end{multline}
Eqs. (21)--(22) do not include radiation energy density, which does not have
influence on the dynamics of inflationary models in the case under consideration
(although, as it was noted above, the contribution
 of ultrarelativistic matter to energy density
can be essentially greater in comparison with scalar field and other components of
gravitating matter), moreover Eqs. (21)--(22) do not contain the parameter $\beta$.
According to Eq. (21) the Hubble parameter in considered approximation is equal to
\begin{equation}
H_\pm=\frac{\displaystyle -3V'\dot{\phi}\pm
\left|4V-\dot{\phi}^2-\sum\rho_i\left(3w_i-1\right)\right|\,
\sqrt{\frac{2\pi}{3M_p^2}\left[4V-\dot{\phi}^2-\sum\rho_i(3w_i-1)\right]-\frac{k}{R^2}}}{\displaystyle
2\left[2V+\dot{\phi}^2+\frac{1}{4}\sum\rho_i(9w_i^2-1)\right]}\, ,
\end{equation}
and  extreme points of the scale factor are
determined by the following condition
\begin{equation}
\frac{k}{R_0^2}+9\left[\frac{V_0'\dot{\phi}_0}{4V_0-\dot{\phi}^2_0-\sum\rho_{i0}\left(3w_i-1\right)}\right]^2=
\frac{2\pi}{3M_p^2}\left[4V_0-\dot{\phi}^2_0-\sum\rho_{i0}\left(3w_i-1\right)\right].
\end{equation}
From Eq. (22) the time derivative of the Hubble parameter at extreme points is
\begin{multline}
\dot{H}_0=\left\{\frac{\pi}{3M_p^2}\,\left[4V_0-\dot{\phi}_0^2-\sum\rho_{i0}\left(3w_i-1\right)\right]^{2}+
\frac{3}{2}\left(V_0'^2-V_0''\dot{\phi}_0^2\right)\right.\\
\left.+9V_0'^2\dot{\phi}_0^2\left[4V_0-\dot{\phi}_0^2-\sum\rho_{i0}\left(3w_i-1\right)\right]^{-1}\right\}
\left[2V_0+\dot{\phi}_0^2+\frac{1}{4}\sum\rho_{i0}\left(9w_i^2-1\right)\right]^{-1}.
\end{multline}
or according to Eq. (24) we can rewrite the expression $\dot{H}_0$ in the following
form
\begin{multline}
\dot{H}_0=\frac{1}{2}\left\{
\frac{27{V'}_0^2\dot{\phi}_0^2}{4V_0-\dot{\phi}^2_0-\sum\rho_{i0}\left(3w_i-1\right)}
+3\left(V_0'^2-V_0''\dot{\phi}_0^2\right)\right.\\
\left.
+\frac{k}{R_0^2}\left[4V_0-\dot{\phi}^2_0-\sum\rho_{i0}\left(3w_i-1\right)\right]\right\}
\left[2V_0+\dot{\phi}_0^2+\frac{1}{4}\sum\rho_{i0}\left(9w_i^2-1\right)\right]^{-1}.
\end{multline}
Obviously Eqs. (24)--(25) correspond to (17)--(18) in considered approximation.

Now in order to investigate inflationary cosmological models in GTG we will analyze
models including scalar fields with positive potentials\footnote{We do not consider
in present paper negative scalar field potentials \cite{mc35}.} $V>0$,
ultrarelativistic matter and component of gravitating matter with energy density
$\rho_1$ and $w_1>\frac{1}{3}$.

\section{Analysis of inflationary cosmological models in GTG}

At first let us consider models including scalar field and ultrarelativistic matter
studied in Ref.~\cite{mc31}. In this case the space $P$ is reduced to the plane of
variables ($\phi$, $\dot\phi$), bounds $L_\pm$ --- to two curves
$\dot{\phi}=\pm\left(4V+|\beta|^{-1}\right)^{\frac{1}{2}}$ and bounce surfaces ---
to corresponding bounce curves on this plane. As it was shown in Ref.\cite{mc31} the
greatest part of inflationary cosmological solutions in the case under consideration
have regular bouncing character, although singular solutions exist also because of
divergence of particular $L_\pm$--solutions. To analyze regular bouncing solutions
we have to examine bounce curves and $H_{\pm}$--functions defined by Eq.(15) on the
plane $P$. Bounce curves have simple form, if the scale of extremely high energy
densities is much smaller than the Planckian energy density. Then according to (24)
bounce curves do not depend on parameter $\beta$ and bounce curves of flat models
are two curves $B_1$ and $B_2$ determined by equation
\[
4V_0-\dot{\phi}_0^2=3\left(\frac{M_p^2}{2\pi}\,{V_0'}^{2}\,\dot{\phi}_0^2\right)^{\frac{1}{3}}
\]
on the plane $P$, which are situated near bounds $L_{+}$ and $L_{-}$
respectively.\footnote{The neighbourhood of origin of coordinates on the plane $P$
is not considered in this approximation, the behavior of bounce curves near origin
of coordinates was examined in Ref.~\cite{mc34}, where scalar fields superdense
gravitating systems were discussed.} Each of two curves $B_{1,2}$ contains two parts
corresponding to vanishing of $H_{+}$ or $H_{-}$ and denoting by ($B_{1+}$,
$B_{2+}$) and ($B_{1-}$, $B_{2-}$) respectively. If $V'$ is positive (negative) in
quadrants 1 and 4 (2 and 3) on the plane $P$, the bounce will take place in points
of bounce curves $B_{1+}$ and $B_{2+}$ ($B_{1-}$ and $B_{2-}$) in quadrants 1 and 3
(2 and 4) for $H_{+}$-solutions ($H_{-}$-solutions) (see Fig.~1).
\begin{figure}[htb!]
\begin{minipage}{0.48\textwidth}\centering{
\epsfig{file=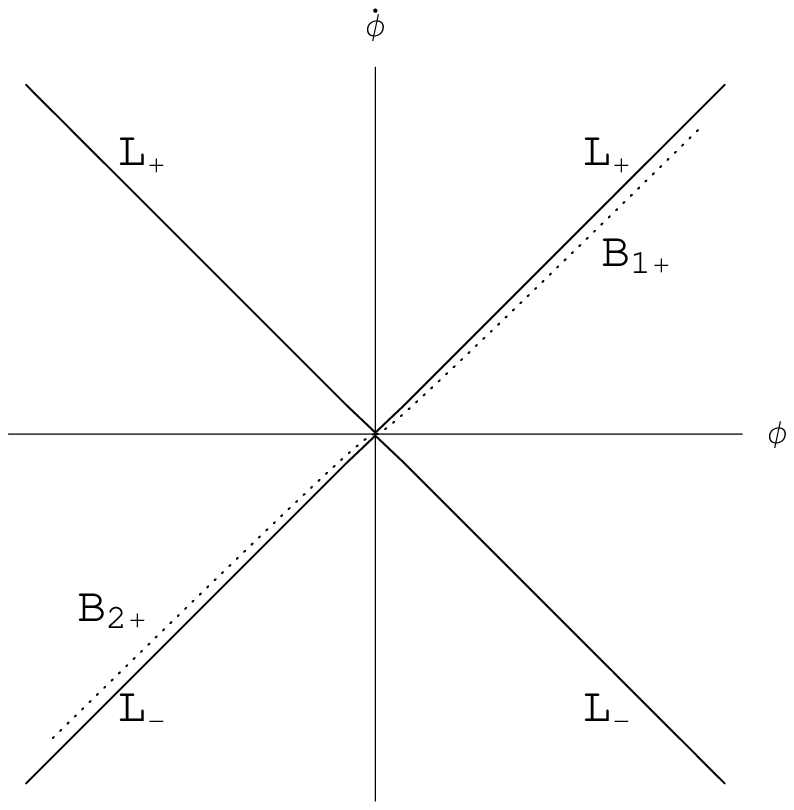,width=\linewidth}}
\end{minipage}\, \hfill\,
\begin{minipage}{0.48\textwidth}\centering{
\epsfig{file=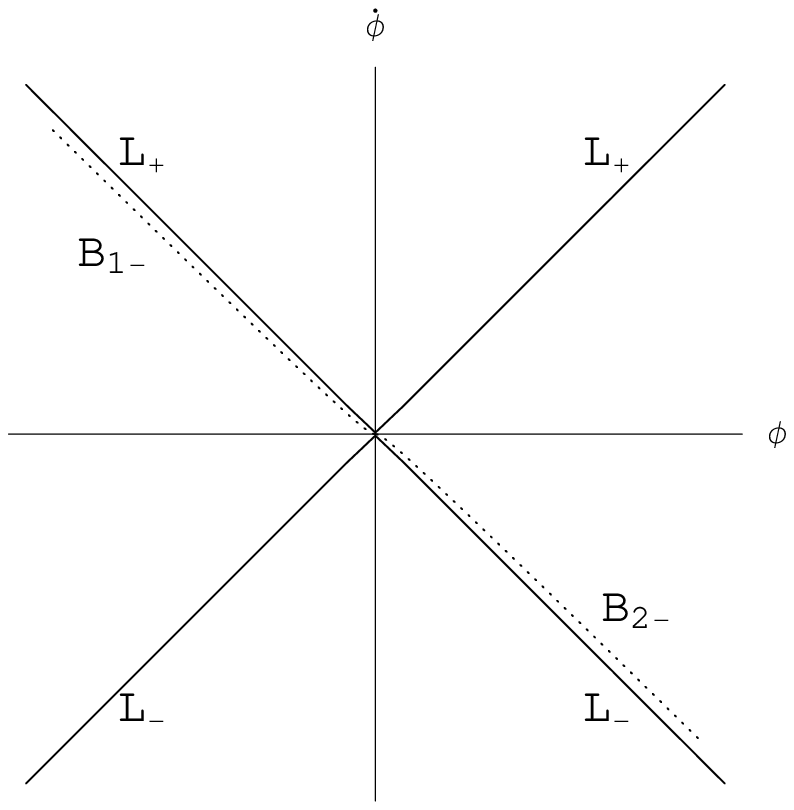,width=\linewidth}}
\end{minipage}
\caption[]{Bounce curves for $H_{+}$-solutions  and
$H_{-}$-solutions in the case of potential
$V=\frac{1}{2}m^2\phi^2$.}
\end{figure}
To analyze flat bouncing models we have to take into account  that besides regions
lying between bounds $L_\pm$ and corresponding bounce curves on the plane $P$ the
sign of values $H_{+}$ and $H_{-}$ for applying potentials is normal: $H_{+}>0$,
$H_{-}<0$. The Hubble parameter $H_{+}$ is negative in regions between curves
($L_{+}$ and $B_{1+}$), ($L_{-}$ and $B_{2+}$), and the value of $H_{-}$ is positive
in regions between curves ($L_{+}$ and $B_{1-}$), ($L_{-}$ and $B_{2-}$). Any
regular cosmological solution has to include both $H_{-}$- and $H_{+}$-solution. The
regular transition from $H_{-}$-solution to $H_{+}$-solution takes place on the
bounds $L_{\pm}$ where $H_{+}=H_{-}$.  If $H_{+}$- and $H_{-}$-solution have with
$L_{\pm}$-bounds one common point, corresponding bouncing solution is regular in
metrics, the Hubble parameter $H$ and its time derivative. If $H_{+}$- and
$H_{-}$-solution are glued with particular $L_{+}$- or $L_{-}$-solution in different
points of bounce, corresponding bouncing solution is regular in $R(t)$ and $H(t)$,
but the derivative $\dot{H}$ in points of gluing has a finite jump. In the case of
open and closed models Eq. (24) ($\rho_i=0$, with the exception $\rho_r\neq 0$)
determines 1-parametric family of bounce curves with parameter $R_0$. Bounce curves
of closed models are situated on the plane $P$ in region between two bounce curves
$B_{1}$ and $B_{2}$ of flat models, and in the case of open models bounce curves are
situated in two regions between the curves: $L_{+}$ and $B_{1}$, $L_{-}$ and
$B_{2}$. Because the behaviour of bounce curves for open models is like to that for
flat models, the situation concerning bouncing inflationary solutions in the case of
open models is the same as described above situation for flat models. Unlike flat
and open models, for which $H_{+}= H_{-}$ only in points of bounds $L_{\pm}$ and
regular inflationary models can be built if $H_{+}$-- and $H_{-}$--solutions reach
bounds $L_{\pm}$, in the case of closed models the regular transition from
$H_{-}$-solution to $H_{+}$-solution is possible without reaching the bounds
$L_{\pm}$. It is because by certain value of $R$ according to (16) we have $H_{+}=
H_{-}$ in the case, if $Z\neq 0$. Such models are regular also in torsion and/or
nonmetricity. Regular inflationary solution of such type was considered in
Ref.\cite{mc33}.

 However, in discussed case there are
also singular solutions. At first note that particular $L_{\pm}$--solutions are
singular. Because bounds $L_{\pm}$ for applying scalar field potentials tend to
infinity, we have that scalar field satisfying on the bounds the equation
$\ddot{\phi}=2V'$ diverges in the past (quadrants 2 and 4) and in the future
(quadrants 1 and 3). In consequence of this any solution including $H_{+}$-solution
(or $H_{-}$--solution) glued with one of the bounds $L_{\pm}$ is singular. Hence the
problem of excluding of singular solutions is connected with regularization of
particular $L_{\pm}$--solutions.

In general case, when approximation (20) is not valid, bounce curves of cosmological
models including scalar fields and ultrarelativistic matter determined by Eq.(17)
depend on parameter $\beta$ . By certain value of $\beta$ we have 1-parametric
family of bounce curves with parameter $\rho_{r0}$ for flat models, and we have
2-parametric families of bounce curves for closed and open models with parameters
$R_0$ and $\rho_{r0}$. The situation concerning cosmological solutions of Eqs.
(9)--(10) does not change.

Now let us analyze models including gravitating matter with energy density $\rho_1$
and $w_1>\frac{1}{3}$ besides scalar field and ultrarelativistic matter. The bounds
$L_{\pm}$ in this case depend on energy density $\rho_1$ and they change by
evolution of models
\begin{equation}
\dot{\phi}=\pm\sqrt{4V+\left|\beta\right|^{-1}-\rho_1\left(3w_1-1\right)\,}\,.
\end{equation}
The Hubble parameter on bounds $L_{\pm}$ according to (14) is
equal to
\begin{equation}
  H=\frac{3\beta V'\dot{\phi}}{1-2 \beta(2V+\dot{\phi}^2)-\frac{1}{2}\beta\rho_1(9w_1^2-1)}.
\end{equation}
Then the equation of scalar field (6) on bounds $L_{\pm}$ because of (27)--(28)
takes the following form
\begin{equation}
\ddot{\phi}+\frac{\displaystyle V'\left(1-4\beta V+
\frac{5+w_1}{1+w_1}\,\beta\dot{\phi}^2\right)}{\displaystyle 1-4\beta V-\frac{1-w_1}{1+w_1}
\beta\dot{\phi}^2}=0.
\end{equation}
If $\frac{1}{3}<w_1\le 1$ and $\beta<0$, the denominator of a fraction in (29) is
positive, and for various potentials $V$ applying in inflationary cosmology (in
particular, $V=\frac{1}{2}m^2\phi^2$, $V=\frac{1}{4}\lambda\phi^4$ etc.) Eq.~(29)
describes finite variations of scalar field between positive maximum and negative
minimum values of $\phi$ at $\dot{\phi}=0$ (see Fig.~2). This means that particular
solution of GCFE becomes regular, and it has bouncing character; a bounce takes
place in points where $\dot{\phi}=0$ (according to (14) and (19) we have in these
points $H=0$ and $\dot{H}>0$). As result, all inflationary cosmological solutions of
GCFE are regular. The analysis of inflationary solutions reaching the bounds
$L_{\pm}$ by numerical integration of Eqs. (11) and (6) is difficult, because the
coefficient at $\dot{H}$ in Eq. (11) tends to zero at $Z\to 0$. Note that bouncing
character have solutions not only in classical region, where scalar field potential,
kinetic energy density of scalar field and energy density of gravitating matter do
not exceed the Planckian energy density, but also in regions, where classical
restrictions are not fulfilled and according to accepted opinion quantum
gravitational effects can be essential.

\begin{figure}[htb!]
\centering{
\begin{minipage}{0.65\textwidth}\centering{
\epsfig{file=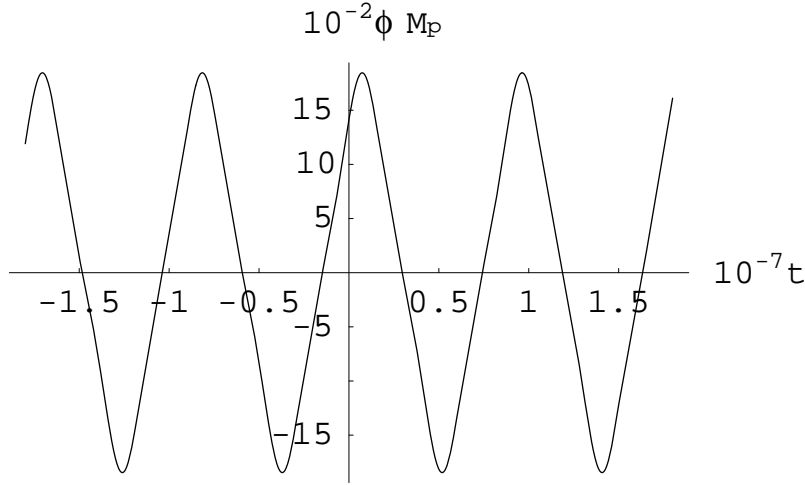,width=\linewidth}}
\caption[]{Numerical solution of Eq. (29) for $V=\frac{1}{2}m^2\phi^2$ ($m=10^{-6}M_p$,
$w=1$, $\beta=-10^6M_p^{-4}$).}
\end{minipage} }
\end{figure}

As illustration of obtained results we will consider particular bouncing
cosmological inflationary solution for flat model by using scalar field potential in
the form $V=\frac{1}{2}m^2\phi^2$ (${m=10^{-6}M_p}$). The solution was obtained by
numerical integration of Eqs. (6), (10) and by choosing in accordance with Eq.(17)
(or (24)) the following initial conditions at a bounce: $\phi_0=\sqrt{2}\, 10^3\,
M_p$, $\dot{\phi}_0=10^{-3} M_p^2$, $\rho_{10}=1.4999\cdot 10^{-6}M_p^4$ ($w=1$,
$\beta=-10^{14}M_p^{-4}$); in accordance with (8) the formula
$\rho_1(t)=\rho_{10}\frac{\textstyle R_0^6}{\textstyle R^6(t)}$ was used, initial
value of $R_0$ can be arbitrary. A bouncing solution includes: quasi-de-Sitter stage
of compression, the stage of transition from compression to expansion,
quasi-de-Sitter inflationary stage, stage after inflation. The dynamics of the
Hubble parameter and scalar field is presented for different stages of obtained
bouncing solution in Figures 3--5 (by choosing $M_p=1$).
\begin{figure}[htb!]
\begin{minipage}{0.48\textwidth}\centering{
\epsfig{file=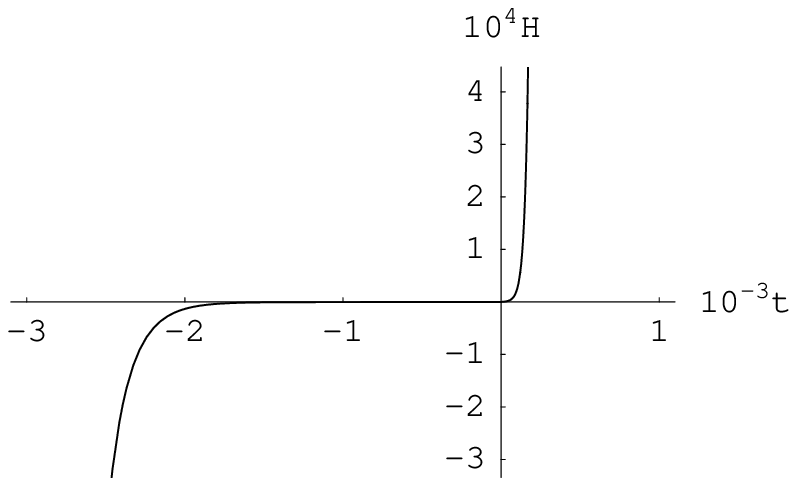,width=\linewidth}}
\end{minipage}\, \hfill\,
\begin{minipage}{0.48\textwidth}\centering{
\epsfig{file=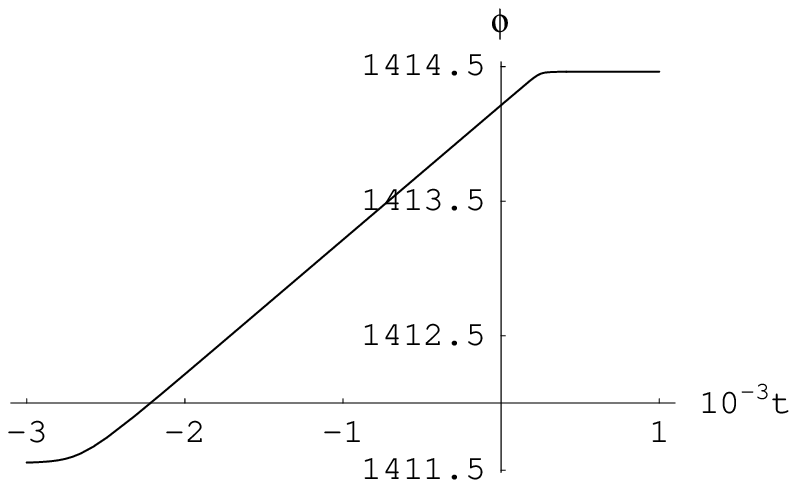,width=\linewidth}}
\end{minipage}\\
\centering{
\begin{minipage}{0.48\textwidth}\centering{
\epsfig{file=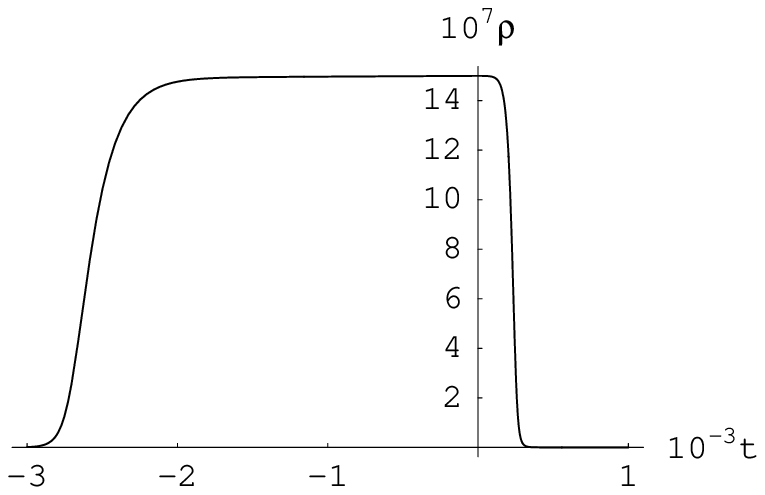,width=\linewidth}}\,\hfill\,
\end{minipage}} \caption{The stage of transition from compression to expansion.}
\end{figure}
\begin{figure}[htb!]
\begin{minipage}{0.48\textwidth}\centering{
\epsfig{file=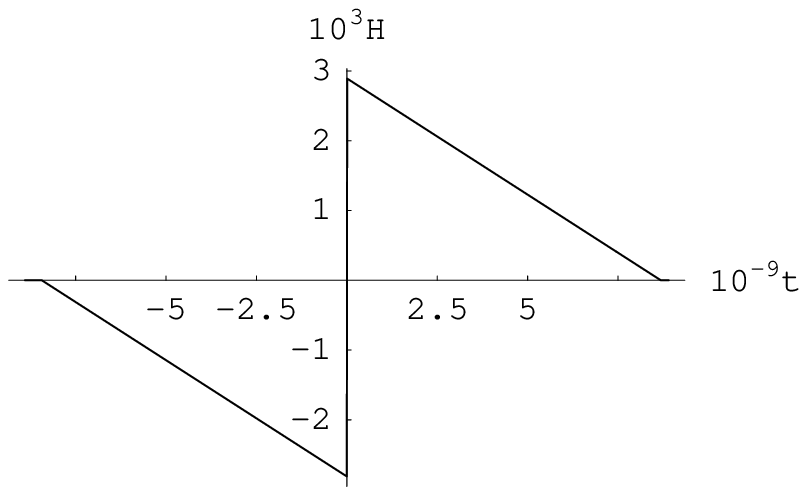,width=\linewidth}}
\end{minipage}\, \hfill\,
\begin{minipage}{0.48\textwidth}\centering{
\epsfig{file=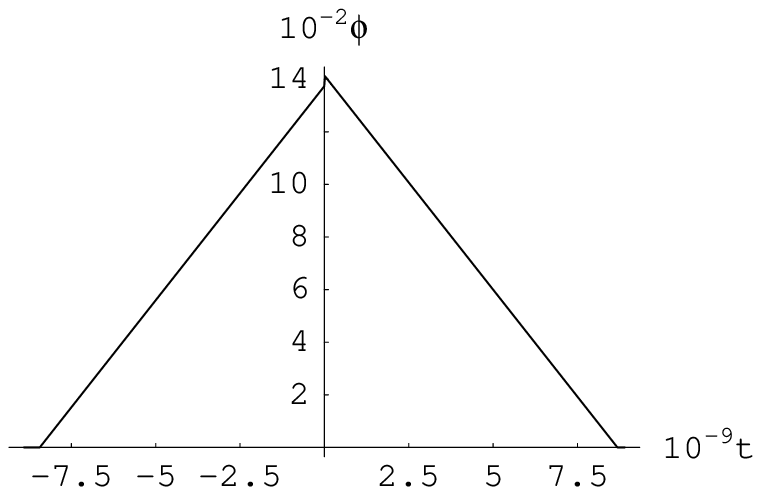,width=\linewidth}}
\end{minipage}
\caption{Quasi-de-Sitter stage of compression and inflationary stage.}
\end{figure}
The transition stage from compression to expansion (Fig. 3) is essentially
asymmetric with respect to the point $t=0$ because of $\dot{\phi}_0\neq 0$. In
course of transition stage the Hubble parameter changes from maximum in module
negative value at the end of compression stage to maximum positive value at the
beginning of expansion stage. The scalar field changes linearly at transition stage,
the derivative $\dot\phi$ grows at first from positive value  $\dot{\phi}_1\sim
1.6\cdot 10^{-7}$  to maximum value $\dot{\phi}\sim\dot{\phi}_0$ and then decreases
to negative value $\dot{\phi}_2\sim -1.6\cdot 10^{-7}$. Quasi-de-Sitter inflationary
stage and quasi-de-Sitter compression stage are presented in Fig.~4. Although the
GCFE (9)--(10) and their approximation (21)--(22) have different structure from
cosmological Friedmann equations of GR, like GR the time dependence of functions
$H(t)$ and $\phi(t)$ at compression  and inflationary stages is linear. The
amplitude and frequency of oscillating scalar field after inflation (Fig. 5) are
different than that of GR, this means that approximation of small energy densities
$\left|\beta\left(4V-\dot{\phi}^2-2\rho_1\right)\right|\ll 1$ at the beginning of
this stage is not valid; however, the approximation (21)--(22) is not valid also
because of dependence on parameter $\beta$  of oscillations characteristics, namely,
amplitude and frequency of scalar field oscillations decrease by increasing of
$|\beta|$ \cite{mc33}. The behaviour of the Hubble parameter after inflation is also
noneinsteinian, at first the Hubble parameter oscillates near the value $H=0$, and
later the Hubble parameter becomes positive and decreases with the time like in GR.
Before quasi-de-Sitter compression stage there are also oscillations of the Hubble
parameter and scalar field not presented in Figures 3--5. Ultrarelativistic matter,
which could dominate at a bounce  as well as another component of gravitating matter
have negligibly small energy densities at quasi-de Sitter stages. At the same time
the gravitating matter could be at compression stage in more realistic bouncing
models, and scalar fields could appear only at certain stage of cosmological
compression. As it follows from our consideration regular character of such
inflationary cosmological models has to be ensured by cosmological equations of GTG.
\begin{figure}[htb!]
\begin{minipage}{0.48\textwidth}\centering{
\epsfig{file=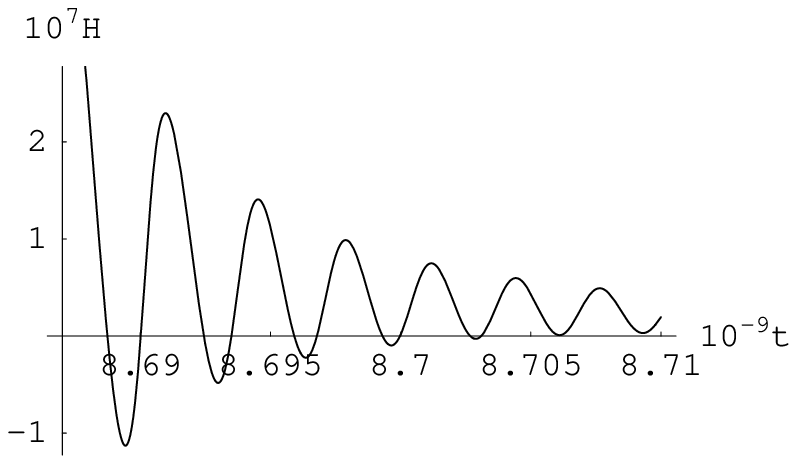,width=\linewidth}}
\end{minipage}\, \hfill\,
\begin{minipage}{0.48\textwidth}\centering{
\epsfig{file=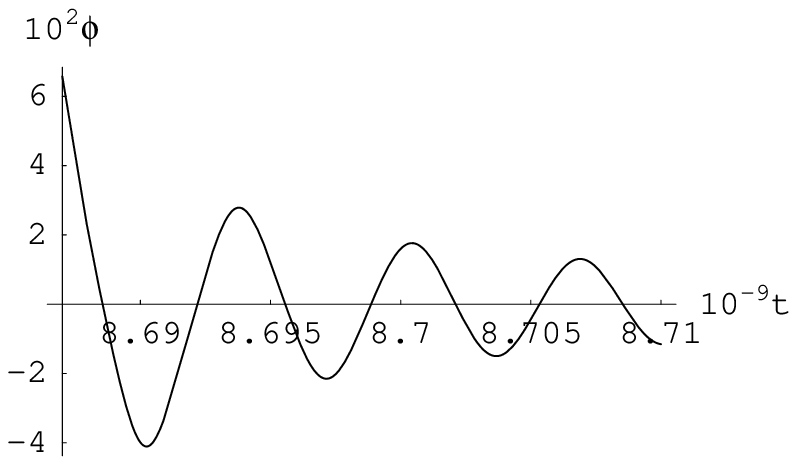,width=\linewidth}}
\end{minipage}
\caption{The stage after inflation.}
\end{figure}

\section{Conclusion}

As it is shown in our paper, GTG  permit to build regular inflationary cosmology, if
gravitating matter at extreme conditions satisfies the following restriction
$\frac{1}{3}\rho<p\le\rho$ and indefinite parameter $\beta$ is negative. All
inflationary cosmological solutions for flat, open and closed models are regular in
metrics and the Hubble parameter. The presence of scalar fields leading to
appearance of inflation in cosmological models changes essentially the structure of
GCFE, as result a family of closed models regular in metrics, the Hubble parameter
and torsion and/or nonmetricity appears. To build realistic inflationary
cosmological models not limited in the time we have to know the change of equation
of state of gravitating matter by evolution of the Universe.

\section*{Acknowledgements}

I am grateful to  Alexander Garkun and  Andrey Minkevich for  help by preparation of
this paper.

\end{document}